\documentclass[universe,article,accept,moreauthors,pdftex]{Definitions/mdpi} 

\firstpage{1} 
\pubvolume{00}
\issuenum{1}
\articlenumber{5}
\pubyear{2021}
\copyrightyear{2021}
\externaleditor{Academic Editor: {Roberto Mignani}} 
\datereceived{19 March 2021} 
\dateaccepted{20 April 2021} 
\datepublished{} 
\hreflink{https://doi.org/} 

\usepackage{amssymb}
\usepackage{CJKutf8}

\Title{Radio Stars of the SKA}
\TitleCitation{Radio Stars of the SKA}

\Author{Bin Yu 
 $^{1,2}$\orcidA{}, Albert Zijlstra $^{2,*}$\orcidB{} and Biwei Jiang $^{1}$\orcidC{}}

\AuthorNames{Bin Yu , Albert Zijlstra and Biwei Jiang}
\AuthorCitation{Yu, B.; Zijlstra, A.; Jiang, B.}
\address{%
$^{1}$ \quad Department of Astronomy, Beijing Normal University, 
Beijing 100875, China; 
  yubin2015g@mail.bnu.edu.cn (B.Y.); \linebreak bjiang@bnu.edu.cn (B.J.)\\
$^{2}$ \quad Department of Physics and Astronomy, The University of Manchester, Manchester M13 9PL, UK} 

\corres{Correspondence: albert.zijlstra@manchester.ac.uk}

\abstract{Radio emission from stars can be used, for example, to study ionized winds or stellar flares. The radio emission is faint and studies have been limited to few objects. The Square Kilometer Array (SKA)  brings a survey ability to the topic of radio stars. In this paper we investigate what the SKA can detect, and what sensitivity will be required for deep surveys of the stellar Milky Way. We focus on the radio emission from OB stars, Be stars, flares from M dwarfs, and Ultra Compact HII regions. The stellar distribution in the Milky Way is simulated using the Besançon model, and various relations are used to predict their radio flux. We find that the full SKA will easily detect all UltraCompact HII regions. At the limit of 10 nJy at 5~GHz, the SKA can detect 1500 Be stars  and 50 OB stars per square degree, out to several kpc. It can also detect flares from 4500 M dwarfs per square degree. At 100 nJy, the numbers become about 8 times smaller. SKA surveys of the Galactic plane should be designed for high sensitivity. Deep imaging should consider the significant number of faint flares in the field, even outside the plane of the Milky Way. }

\begin{document}



\section{Introduction}
In the past few decades, improvements in the sensitivity of instrument such as the Very Large Array (VLA) have allowed the study of radio emissions from stars.  VLA observations have shown that radio emissions from stars are more common than was expected based on the Sun's radio emission and the detection possibility is higher than had been assumed~\cite{Gudel2002,White2004}.  Radio emission has been detected from many different kinds of stars in different evolution stages, including mass-losing stars and binaries with thermal wind, magnetically-active stars, and other main sequence stars~\cite{White2004}. Some radio emissions from these sources may be produced by similar processes occuring on the Sun, but not all of them can be explained in this way. Since it is the atmosphere that produces the radio emission, studying the radio emission can help us understand the properties of the star's atmosphere and the environments in which those astrophysical phenomenon and stellar activities are taking place.

Depending on the environments, several kinds of radio emission mechanisms are operating in the different parts of the stellar atmosphere.  Thermal free-free or bremsstrahlung emission mostly comes from stellar outflows and chromospheres, and this mechanism is responsible for the classic stellar-wind radio emission. For stars with solar-like magnetic activity, non-thermal synchrotron emission is generated in the flares. Depending on the magnetic field strength, this mechanism produces gyrosynchrotron emission for mildly relativistic electrons and synchrotron emission for highly relativistic electrons. For stars with radio brightness temperature higher than $10^{12}$ K, there are two main kinds of coherent radiation mechanisms~\cite{Gudel2002}. The first is plasma radiation. It allows one to approximately determine the electron density in the source. Its fundamental emission is best observed below 1~GHz. The second is the electron cyclotron maser emission~\cite{MelroseDulk1982,MelroseDulk1984}. It is emitted mostly at the fundamental and the second harmonic of $\nu_c$ (\ref{eq:maser}).

\begin{equation}
    \nu_{c} \equiv \frac{\Omega_{c}}{2 \pi}=\frac{e B}{2 \pi m_{e} c} \approx 2.8 \times 10^{6} B\, [\mathrm{Hz}]
\label{eq:maser}
\end{equation}
where $c$ is the speed of light, and the magnetic field strength $B$ is given in Gauss.  This can therefore be used to determine the magnetic field strength in the source.

However, the detection of radio stars remains limited by the sensitivity of current telescopes. The number of detected radio stars is not enough to build a complete sample, causing a bias. The bias comes from the situation that the observation ability we have for now only allows us to detect stars with the highest radio luminosity or stars that are located at a very close distance. To trace lower radio luminosity, a next generation radio telescope is needed. The Square Kilometer Array (SKA) will be an interferometric array of antenna stations located in different places of the world, building up a total collecting area of approximately one square kilometer sometime in the 2020s, making it tens of times more sensitive than any other radio instrument. The original aim was to make it possible to detect radio emission as weak as 10 nanoJy in 8-h integration time~\cite{Richard2004}. This significant improvement will dramatically increase the observable number of radio stars and bring about a new window of the radio universe. The discovery of a new-type of stars and phenomena never seen before is highly possible.

The work in~\cite{White2004} discussed some likely advances that SKA would bring us in stellar radio astronomy. We briefly summarize it here. Due to the improved sensitivity, a wide range of main-sequence stars other than the Sun will be detectable at radio wavelength, allowing us to build a comprehensive sample for study. For cool main-sequence stars, VLA observations found that they may have a non-thermal corona emitting strong and steady radio emission. The SKA will likely help us reveal the source of it. With a better spatial resolution, SKA can produce much better radio imaging of the stellar atmosphere than we now have. For example, from 4.6 to 8.5~GHz, SKA 1 is expected to have a resolution of 0.08 arcsec. In general, the resolution of SKA1 will be 4 times better than the JVLA~\cite{Braun2019}. The  resolution of the full SKA will improve significantly on the base of the SKA1. Along with the spectra of  thermal atmospheres, SKA will provide constraints in building the stellar atmosphere model. For pre-main sequence sources, objects in star-forming regions over 100 pc away will become detectable, especially for those fainter sources that do not look like nearby flare stars and active binaries. For PlanetaryNebulae (PNe), the sensitivity of the SKA and its fast survey speed will enable the detection of every PN in the Milky Way, showing us the missing population of PNe ~\cite{Umana2015aska}.The radio detections of stellar outflows often are very faint and the spectral information is also limited. The SKA can improve our understanding of the nature of them and study of the mechanism ongoing. The flare phenomena shows information of energy releases in the atmospheres of stars. With the SKA, we will be able to observe radio flares and monitor its variation for stars of various types. A large sample can tell us the relation between flares and stellar properties such as ages and masses, providing a better understanding of the physics in stars. 
For now, our knowledge about stellar winds mostly comes from the hot stars, since the emission from the winds of cool stars is very weak and hard to detect. The SKA may improve this situation and give us a clearer view of it. All the stellar radio astronomy questions mentioned above are but a small fraction of the questions that can be answered by the SKA.

In hot stars, the classic stellar-wind radio emission is produced by the thermal free-free emission. In addition, non-thermal synchrotron emissions may also exist. The physics process producing it is not totally clear yet. More wind detections from the SKA towards various types of stars will enrich our research samples in this field. In conclusion, the improvements in both sensitivity and spatial resolution of SKA will significantly contribute to our study of stellar radio astronomy in almost every aspect.

To study properties of a wide range of stellar types, the Hertzsprung--Russell diagram is a very useful tool. A radio HR diagram is presented as Figure 1 
in~\cite{Gudel2002}, based on stellar radio detections between 1 and 10~GHz of 440 stars. 
The survey from SKA will construct a far more comprehensive radio stars sample for building the radio HR diagram.

The unparalleled sensitivity of SKA traces an unexplored region of parameter space. This makes it hard to predict the actual detections, especially for the Galactic sources considering the diversity of the radio emission contributors. A model of the stellar radio emission of the Milky Way down to nano-Jy levels is of great value for future SKA observation. We introduce the Besançon computer model and simulate the distribution and properties of stars in the galaxy. After applying the radio emission mechanisms for each type of stars, a simulated HR diagram of the galaxy corresponding to the sensitivity of SKA can be built.

In this work, we focus on the radio emission from OB stars, Be stars, and Ultra Compact HII regions  for massive stars, and for M dwarfs for low-mass stars. The paper will predict the source density and brightness distributions which the SKA may detect for these types of stars in different directions within the Milky Way.

\section{Materials and Methods}
\subsection{The Besançon Model}
Based on comprehensive analysis about the theories and models of the formation and evolution of the galaxy, the stars, the stellar atmosphere, and dynamical constraints from observations, the Besançon model produces a  picture of the galactic populations consistent with the scenario of galaxy evolution. It predicts the distribution of stars at each location in the galaxy, based on the thin disk, the thick disk, the halo, and the bulge.

With given parameter inputs, the model is able to predict the probable properties of star samples, including the distance, age, type, effective temperature, absolute magnitude, apparent magnitudes, and colors. It integrates along a line of sight. The user can select on each of these parameters to limit the output to a particular set of stars. This is useful to prepare observations such as the future SKA survey. Details about the theories and models adopted by the Besançon model can be found in the original paper~\cite{Robin2003}, and followed by some revisions enabled by the latest study, such as the bar population~\cite{Robin2012A&A}, the thick disc and halo populations~\cite{Robin2014A&A}, new kinematics~\cite{Bienaym2015A&A,Robin2017A&A}, and the outer disc scale lengths, warp, and flare by~\cite{Amores2017}.

In practice, the model contains a description of the distribution of stars of a different mass and evolutionary state over the galaxy. For a line of sight and field size, the models produce the number of such stars as a function of distance. The model is gridded to a distance step size specified by the user and returns a table of stars with their parameter bases on the input selection provided by the user. 

In this work, the default version (m1612) of the Besançon model is adopted together with all the other default configurations, such as the Johnson-cousins photometric system. 

To obtain an adequate sample, several constraints are applied in the use of the Besançon model, as listed below.
 
 For OB stars, the distance range of modeled stars is set from 0 to 30 kpc to the sun, with a distance step of 50 pc. The sky coverage is the whole galactic plane ($-1^\circ\leq b \leq1^\circ ,$\linebreak$ 0\leq l \leq360^\circ $), with sampling steps of $0.2^\circ$ in latitude and $1^\circ$ in longitude, amounting to 720 square degrees. Apparent magnitudes in the V band are set from 0 to 18, for no extinction. (The interstellar extinction was set to zero in our Besançon model, since radio emission is not affected by it and is calculated from the absolute magnitudes and distance.) Spectral types from O3 to B9 of all luminosity classes and age groups are included. Since only the density distribution is needed in this work, the kinematics are not included. Using the inputs described above, we generated a sample of 6.33~million OB stars. 
 
For the M dwarfs, the latitude range is expanded to $-20\leq b \leq20^\circ$ and because the M dwarfs have higher scale height and lower luminosity, the observed distribution is much less focussed to the plane. The longitude range is limited to $0\leq l \leq10^\circ$, because of the large amount of M dwarfs which would otherwise exceed our calculation ability. The total area is 400 square degrees. Since we expect to mainly see nearby M dwarfs, the longitude distribution is expected to be less important. Spectral types from M0 to M9 of all luminosity classes and age groups are included. We take the $10^\circ\times40^\circ$ region towards the galactic center as an example that contains 68.18 million M dwarfs.

Interstellar extinction is included in the model. However, for radio emission the optical extinction should be ignored. We will use the bolometric magnitude and distance. Basic stellar spectral types are included. The Besançon model does not include the galactic centre region and it does not cover all possible stellar classes. 

As an example, the HR diagram of the sample of OB stars and M dwarfs used in the current paper (see below)  is presented in Figure \ref{fig:HR}. It shows the hot end of the main sequence. The temperature and magnitude resolution of the Besançon model is visible in the plot. A few evolved blue supergiants are included in the original sample. They only account for an extremely small fraction in the whole sample. Supergiants are not covered by the radio emission relations for OB stars, and their emission mechanism may differ from that of main sequence stars.  Therefore, they are not considered in this~paper.

\begin{figure}[H]
\includegraphics[width=12cm]{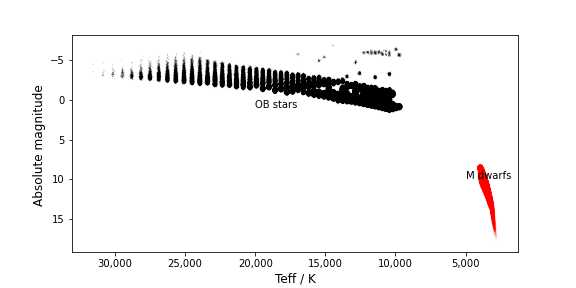}
\caption{The HR diagram of the sample generated from the Besançon model.}
\label{fig:HR}
\end{figure}   

The Besançon model includes all spectral types and luminosity classes, but not all stages of stellar evolution. The Padova isochrones are less accurate for the AGB phase, and terminate at the start of the TP-AGB. Post-AGB evolution is not included and white dwarfs are added afterwards. Binary evolution is not covered. Important groups that are excluded are symbiotic stars, high-mass binaries, and planetary nebulae.

\subsection{The Square Kilometre Array}
\label{sec:SKA}
The SKA will be a multi-national, next-generation radio telescope. Once completed, it will be the largest scientific instrument in the world with a collecting area of more than one million square meters. This project is divided into two phases (SKA1 and SKA2). SKA1 is currently under construction in Australia and  South Africa, covering the 50 MHz--350~MHz (SKA1-Low) and 350MHz--14GHz (SKA-Mid) range of the spectrum, separately. 

Ref \cite{Braun2019} presents a recent anticipation of SKA1’s performance. The continuum sensitivity of SKA1 can reach from 26 $\upmu$
Jy/beam at 110 MHz to 1.2 $\upmu$Jy/beam at 12.5~GHz, after one hour of integration time. SKA1 is estimated to be in use by 2023.

SKA1 only constitutes about 10$\%$ of the full array. SKA2 is planned to start after the completion of SKA1. The number of arrays will be extended to 2000 and the frequency range will expand towards higher frequencies (up to 25~GHz). The point source sensitivity will approach 10 nano-Jy in 8~h of integration~\cite{Combes2015}.

In this work, we will assume observations at 5~GHz. The frequency of 5~GHz is chosen for the typically steep thermal spectra of stellar radio emission. The emission increases steeply with a higher frequency. It also benefits from lower confusion with non-thermal emission, giving better sensitivity.

{\bf At our chosen frequency of 5~GHz}, the SKA1 is expected to achieve a point source sensitivity of 1.4 $\upmu$Jy in 1 h (5$\sigma$). This assumes a bandwidth of 1.5~GHz ($\Delta \nu/\nu \approx 0.3$) and is appropriate for synthesized beams between 0.17$''$ to 23$''$. The full bandwidth available is 3.9~GHz, which improves the sensitivity for point sources by a factor of 1.6. In 8~h of integration at 5~GHz, the SKA1 can reach 300 nJy.  Natural $uv$ weighting in sparse fields (allowing the shortest baselines to be used) can provide another factor of 2 improvement but this may be difficult in the galactic plane, even at 5~GHz where extended synchroton emission is minimized.

SKA2 will be an order of magnitude more sensitive. We will here assume that it can indeed reach 10 nJy in 8~h. This is the most optimistic scenario and assumes perfect performance, full usable bandwidth, ideal conditions, and empty fields.

\subsection{The Radio Emission of OB Stars}
As early-type massive stars, OB stars commonly undergo mass loss by a stellar wind. Their radio emission is believed to be generated mainly by the thermal free-free emission from circumstellar ionized gas, which is described by the classical line-driven stellar wind theories~\cite{Wright1975,Lamers1999,Vink2011}. In these stars, Fe III and IV ions are accelerated by radiation then interact with other atoms in the photosphere. The free–free radio flux density $S_\nu$ in Jy is related to the stellar mass-loss rate $\dot{M}$ by~\cite{Wright1975,Panagia1975}:

\begin{equation}
    \dot{M}=\frac{3.01 \times 10^{-6} \upmu}{Z\left(\gamma g_{\mathrm{ff}} \nu\right)^{1 / 2}} V_{\infty} S_{\nu}^{3 / 4} D^{3 / 2}\, M_{\odot} \mathrm{yr}^{-1} ,
\label{eq:mloss}
\end{equation}
where $\upmu$ is the mean ionic weight (in amu), $V_\infty$ is the terminal velocity in $\rm km \,s^{-1}$, $D$ is the distance to the star in kpc, $\nu$ is the frequency in~GHz, $\gamma$ is the mean number of electrons per ion, and $Z$  is the rms charge per ion. $g_{\mathrm{ff}}$ is the free-free Gaunt factor given by:

\begin{equation}
    g_{\mathrm{ff}}=-1.66+1.27 \log \left(T_{\mathrm{wind}}^{3 / 2} /(Z \nu)\right)
\label{eq:gaunt}
\end{equation}
where $T_{\rm wind}$ is the local temperature (in K) at the radio photosphere.

We utilize Equation \eqref{eq:mloss} to estimate the radio flux of our sample OB stars simulated by the Besançon model. The winds of all stars are assumed  to be fully ionized and helium is assumed to be singly ionized, which gives $Z=\gamma=1$.  $\upmu  =1.26$ is taken as a uniform mean ionic weight. As a compromise of the sensitivity and survey speed, $\nu =5$\,GHz (6 cm) is chosen as our survey frequency. Distances are taken from the catalogue generated by the Besançon model. $T_{\rm wind}$ is adopted as $0.5T_{\rm eff}$, following ~\cite{Bieging1989}. 

To calculate the theoretical mass-loss rates of the sample stars, we adopted the method from~\cite{Vink2011}, which is based on the  Monte--Carlo approach developed by~\cite{Abbott1985}, and multiple scatterings are taken into account. The mass loss recipe presented by~\cite{Vink2000A&A...362..295V} is separated into two parts around 25,000 K by the so-called ``bi-stability jump'', which is caused by the abrupt change of driving lines from iron ions. On the hotter side between 27,500 
 and 50,000~K, the formula is given by:
\begin{equation}
    \begin{array}{ll}
\log \dot{M}=&-6.697(\pm 0.061) \\
&+2.194(\pm 0.021) \log \left(L_{*} / 10^{5}\right) \\
&-1.313(\pm 0.046) \log \left(M_{*} / 30\right) \\
&-1.226(\pm 0.037) \log \left(\frac{v_{\infty} / v_{\mathrm{esc}}}{2.0}\right) \\
&+0.933(\pm 0.064) \log \left(T_{\mathrm{eff}} / 40,000\right) \\
&-10.92(\pm 0.90)\left\{\log \left(T_{\mathrm{eff}} / 40,000\right)\right\}^{2} \\
 \multicolumn{2}{l}{\text { for } 27,500 < T_{\mathrm{eff}} \leq 50,000 \mathrm{K}}
\end{array}
\label{eq:ml1}
\end{equation}
where $\dot{M}$ is in $M_{\odot}\, \rm yr^{-1}$, $L_*$ and $M_*$ is in solar units, and $T_{\rm eff}$ is in Kelvin. 

On the cooler side between 12,500 K and 22,500 K, the formula is:
\begin{equation}
    \begin{array}{ll}
\log \dot{M}=&-6.688(\pm 0.080) \\
&+2.210(\pm 0.031) \log \left(L_{*} / 10^{5}\right) \\
&-1.339(\pm 0.068) \log \left(M_{*} / 30\right) \\
&-1.601(\pm 0.055) \log \left(\frac{v_{\infty} / v_{\mathrm{esc}}}{2.0}\right) \\
&+1.07(\pm 0.10) \log \left(T_{\mathrm{eff}} / 20,000\right) \\
 \multicolumn{2}{l}{ \text { for } 12,500<T_{\mathrm{eff}} \leq 22,500 \mathrm{K}}.
\end{array}
\label{eq:ml2}
\end{equation}

The ratio $v_\infty/v_{esc}$ is 2.6 for the hot side of the jump and 1.3 for the cool side of the jump as given by~\cite{Lamers1995}. For an effective temperature between 22,500 K and 27,500 K, the position of the jump needs to be determined specifically. Details can be found in the original paper. They also present a rough estimation of mass-loss rates of stars below 12,500K, for which the ratio $v_\infty/v_{esc}$ drops to 0.7 and the constant in Equation \eqref{eq:ml2} is increased to a value of~$-5.99$. 

\begin{figure}[H]

\includegraphics[width=10cm]{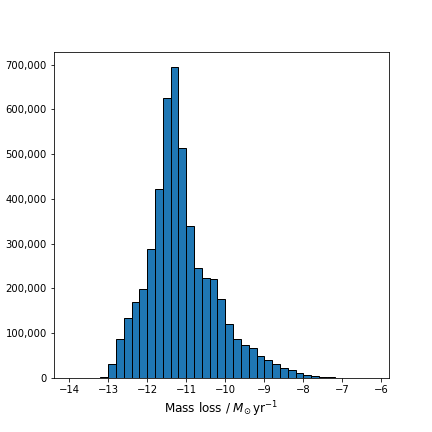}
\caption{Mass-loss rates of our sample OB stars.}
\label{fig:Mlos}
\end{figure}

\textls[-30]{The mass-loss rate level of our OB sample ranges from $10^{-14}\,M_{\odot}\, \rm yr^{-1}$ to $10^{-6}\,M_{\odot}\, \rm yr^{-1}$.} The details can be found in Figure~\ref{fig:Mlos}.

These numbers are for the winds of single stars. A significant fraction of OB-type stars are in binaries~\cite{Sana2012}. The wind interaction between the two stars of the binary can cause a separate radio emission, which may be much brighter than that of the single stars.

\subsection{The Radio Emission of Be Stars}
\label{sec:Be}
By definition, classical Be stars are hot, non-supergiant, non-radially pulsating, rapidly rotating B-type stars, which show or have shown one or more Balmer lines in emission (most prominently in $\mathrm{H}\alpha$)~\cite{1981BeSN....4....9J,Zorec1997A&A...318..443Z,Rivinius2013}. Their rotation speeds are presumed to be near their critical velocity and they have strong stellar winds and a high mass-loss rate. The emission lines are believed to be produced in the surrounding gaseous circumstellar envelope. A viscous, Keplerian, circumstellar “decretion” disk is formed from ejected stellar mass in discrete events. These disks together with the emission lines can be variable on time-scales from hours to decades~\cite{Rivinius2013}. The nature of Be stars and their disks are not yet well understood. Several possible mechanisms are proposed, such as the wind compressed disk model of~\cite{Bjorkman1993}, and the SIMECA code by~\cite{Stee1994,Stee1995}. Despite some debates, the agreement is that the disk strongly correlates with rapid rotation~\cite{Porter2003}. 

Be stars are usually very bright and over-luminous compared to “normal” B stars owing to the active stellar activity and the disk extending several stellar radii along the equatorial plane of the star~\cite{Struve1931,Gies2007,Carciofi2009}, or several hundred to 1000 solar radii~\cite{Taylor1990}. Be stars are very common; they account for about 20$\%$ of all B stars, based on previous observations~\cite{Rivinius2013}. They are non-neglectable radio sources in future galactic plane surveys. The first radio detection of a Be star was the B5 Ve star $\psi$ Per at 6 cm with the VLA~\cite{Taylor1987}, followed by 5 more VLA detections at 2 cm among 22 Infrared Astronomical Satellite (IRAS) Be stars~\cite{Taylor1990}. These measurements yielded significantly steeper spectral indexes ($>$1) than found at an infrared wavelength. While the origin of this steepening of the spectral index remains unclear, the correlation between radio and $\mathrm{H}\alpha$ luminosity suggests that these share the same origin. 

Only a few Be stars were detected by~\cite{Taylor1990} due to the detection sensitivity which was limited to about 0.1 mJy.  With the SKA, the completeness of the Be star samples will be significantly extended. 

Be stars are not specifically included in the Besançon model. In this work, we randomly pick 20$\%$ of the B stars simulated by the Besançon model and assume them to be Be stars, as mentioned above. This will only select main-sequence Be stars. Other Be stars (for example, Herbig Be stars, B[e] stars) are not included. The radio radiation mechanism of Be stars is not clear yet. So we give rough estimates of their radio fluxes based on the luminosity presented by~\cite{Taylor1990}. In Figure 4 
 of~\cite{Taylor1990}, there is a tendency that later spectral types have a lower radio luminosity, where the luminosity drops around type B5. Based on the values in their sample, for stars with the spectral type earlier than B5, a luminosity density at 14\,GHz of $2.5\times10^{16}\, \rm erg\, s^{-1}\, Hz^{-1}$ is assumed, and for the rest we give $0.2\times10^{16}\, \rm erg\, s^{-1}\, Hz^{-1}$. A 50 percent uncertainty is implemented by multiplying each value by a normal distributed random number between 0.5 and 1.5. After transferring to 5\,GHz using a spectral index of 1.3 as observed at $\psi$ Per, the flux at 5~GHz of Be sample stars can be calculated using the distances of the simulated stars with \ref{eq:FL}.

\begin{equation}
    F= \left(\frac{5}{14}\right)^{1.3} 10^{-19} \left(\frac{L}{4 \pi D^{2}}\right)
\label{eq:FL}
\end{equation}
where $L$ is the luminosity density at 14~GHz in erg\,s$^{-1}$\,Hz$^{-1}$, $D$ is the distance in meters, and $F$ is the radio flux density at 5 GHz in Jy.

\subsection{The Radio Emission of Ultra Compact HII Regions }
\textls[-15]{UCHII regions are small photo-ionised nebulae that have not expanded out of the molecular cloud where they were born. Characteristic physical properties are \mbox{diameters $\lesssim0.1$ pc,} electron number densities $\gtrsim10^4\,\rm cm^{-3}$, and emission measures $\gtrsim10^7\,\rm pc\,\rm cm^{-6}$ \cite{Wood1989a}. }

UCHII regions are produced through the ionization of the gas by extreme ultraviolet radiation from a newly-formed high-mass star after the collapse of a dense molecular cloud. This process can emit bright radio free-free emission that can penetrate the natal clouds, while the optical and near-infrared emission is blocked~\cite{Churchwell2002}. Based on observations, their ionizing photon emission rates are about $10^{44}\,\rm s^{-1}$ to$10^{49}\,\rm s^{-1}$ corresponding to B2- to O5-type~stars.

The life length of the UCHII phase is estimated to be about 3--10$\%$ (3--4$\times\,10^5$ yr) 
 of the main sequence OB star lifetime~\cite{Wood1989b,Wood1989a,Mottram2011}. The considerable number fraction and intense radio flux makes them well detectable in the radio wavelength. UCHII regions provide information about the early life of massive stars and their environment. Observations of UCHII regions with higher sensitivity and resolution by SKA can measure star formation rates throughout the galaxy. We present estimates of their radio emission here. The SKA will be able to spatially resolve these objects, but here we only consider the integrated~flux.

Similar to the process in the previous section, 3$\%$ of OB stars in the Besançon sample are assumed to be UCHII regions, based on their lifetime. This will overestimate the numbers for the lower mass B stars that live far longer than the O-type stars. Our work will therefore include the much more extended HII regions for the lower mass~stars.

To estimate the radio flux, first the relation between H$_\beta$ luminosity and the total stellar luminosity in the nebula by~\cite{Zijlstra1989a} (Equation \eqref{eq:L_Hbeta}) is used to calculate the $L_{H_\beta}$ from $L_\star$ (converted from bolometric magnitudes available in the Besançon Model):

\begin{equation}
\frac{L_{H_{\beta}}}{L_{\star}}=h \nu_{H_{\beta}} \frac{\alpha_{H{\beta}}}{\alpha_{B}} \frac{15 G_{1}\left(T_{\star}\right)}{\pi^{4} k T_{\star}} .
\label{eq:L_Hbeta}
\end{equation}

\noindent Here, $h$ and $k$ are the Planck constant and the Boltzmann constant respectively. $\nu_{H_\beta}$ is the H$_\beta$ frequency in Hz. $\alpha_{B}$ is the recombination coefficient for all Balmer transitions together and $\alpha{H_\beta}$ is the coefficient for the H$_\beta$ transition alone. The ratio between them is approximately 0.116. $G_1(T_\star)$ is a function of stellar temperature $T_\star$, and can be found in~\cite{Zijlstra1989b}. After obtaining  $L_{H_\beta}$, with the luminosity-flux relation and the radio to H$_\beta$ flux ratio by~\cite{Pottasch1984}, which is $S_{\nu} / F(\mathrm{H}_\beta)=2.51 \times 10^{7} T_{\mathrm{e}}^{0.53} \nu^{-0.1} Y\left(\mathrm{Jy}\, \mathrm{mW}^{-1}\, \mathrm{m}^{2}\right)$, the flux of the sample UCHII regions at 5 GHz can be calculated. $T_e=10^4 \,\rm K$ is the electron temperature, $\nu$ is the radio frequency in~GHz, and $Y$ is a factor incorporating the ionized He/H ratio.

\subsection{The Radio Emission of M Dwarf Flares}

Stellar flares are highly energetic events caused by magnetic reconnection in the stellar corona~\cite{Haisch1991,Martens1989SoPh..122..263M,Shibata2011LRSP....8....6S}. They usually last minutes to a few hours, and emit energy ranging form $10^{23}$~erg~\cite{Parnell2000ApJ...529..554P} to $10^{33}--10^{38}$
 erg~\cite{Shibayama2013ApJS..209....5S}. In the standard model, the released energies are transported to the lower atmosphere causing bright coronal and chromospheric emissions. Information from these emissions can help us understand what is happening at the reconnection site. Similar to the sun, these emission can range across the electromagnetic spectrum from the radio to the X-ray~\cite{Hawley1995ApJ,Hawley2003ApJ,Berger2008ApJ}. 

Radio bursts from M dwarfs are of primary interest in our study of stellar flares. First, flares are related to magnetic fields, which are common on active M dwarfs. In the work of~\cite{Gunther2020AJ}, more than 40$\%$ of mid- to late-M dwarfs and about 10 $\%$ of the early M dwarfs detected by the TESS (Transiting Exoplanet Survey Satellite) shows observable flares. Second, about 70$\%$ of the stellar population are M dwarfs, providing sufficient potential samples to study~\cite{Henry1994AJ,Reid2004AJ,Covey2008AJ}. Furthermore, as flares are energetic events which can significantly impact the interplanetary environment of the host star, understanding flares is of great importance in the search for habitable exoplanets around M dwarfs. 

While flares from M dwarfs are common, the radio detection of them are limited by the sensitivity,  wavelength range, and other constraints of existing instruments. Current detections are mostly from nearby active M dwarfs~\cite{Romy2020ApJ...892..144R}. The improvement from SKA will lead us to a deeper and much larger sample of flaring M dwarfs and reveal the uniformity and diversity of solar and stellar flares.

To estimate the number and radio flux of flares of M dwarfs, the fraction of active M dwarfs among the whole M dwarf population is needed. A previous study shows that the ratio increases as a function of the spectral type, from a few percent at M0 to nearly 100$\%$ at M9 \cite{Hawley1996AJ,Gizis2000AJ,West2004,West2008AJ}. In this work, we adopt the newer numbers from~\cite{West2008AJ}, which  agree with later results from~\cite{Hilton2009AIPC}. The numbers are read from Figure 3 
 of~\cite{West2008AJ}. From M0 to M9, 0.058, 0.068, 0.100, 0.167, 0.318, 0.661, 0.689, 0.838, 0.986, and 0.902 are set as the fraction of active M dwarfs. Stars are randomly picked based on these proportions. 

The possible mechanisms of flares include coherent emission mechanisms such as plasma emission and electron cyclotron maser emission. However, there still remains a considerable amount of uncertainty, making it challenging to predict their flux precisely. Using observations with the VLA,~\cite{Villadsen2019} presented a ultra-wideband radio burst sample based on five active M dwarfs. A total of 22 bursts were detected. The detected fluxes for each time sequence are given in their work. They observed for 40 h in total, divided over 5 stars, giving 8~h on average per star. The flux densities are averaged over 10 min of integration. 

Based on this, a probability density function is calculated with a Gaussian kernel density estimation method for the flux of the bursts at 2.8--4~GHz (Figure \ref{fig:MFlux}), for a single integration time of 10 min targeted at an active star. This distribution is applied to the stars randomly selected from the M dwarf catalogue generated by the Besançon model. The data from~\cite{Villadsen2019} shows that flares decay over time scales that are longer than 10 min. We therefore assume that each flare lasts 20 min. As a result, the integrated flux is doubled from the estimated value for 10 min of integration. For one 20-min flare, the detection limit for the flare is about 5 times worse (50 nJy) compared to the nominal 8-h sensitivity. 

\begin{figure}[H]
\includegraphics[width=10cm]{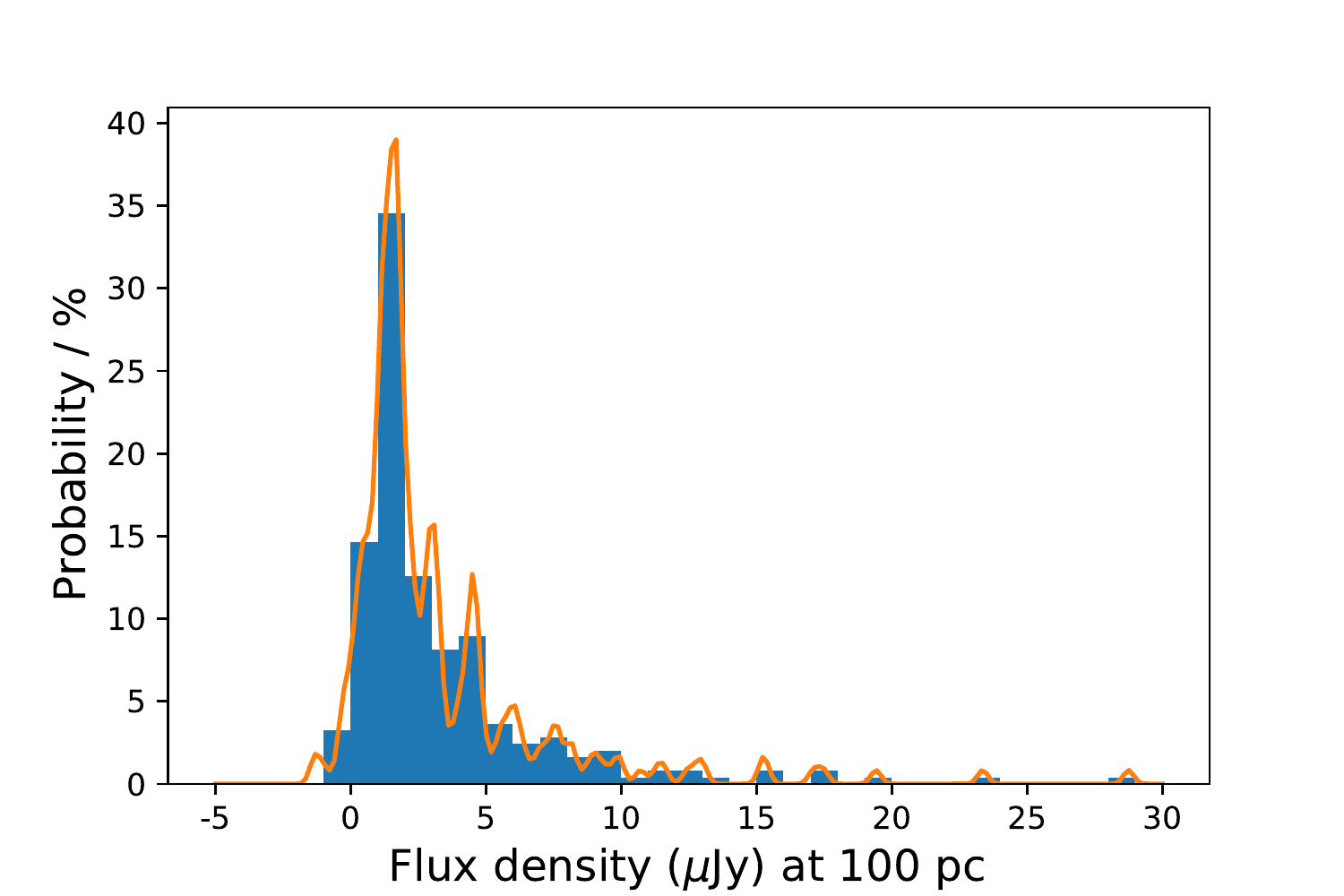}
\caption{Flux density distribution of the burst sample. The histogram shows the probability density distribution of the 10-minute flux density at 1 pc at 2.8-4~GHz, from the results of~\citet{Villadsen2019}. The probability density function generated for the data is shown as the curve.}
\end{figure}   

The flux probability density function we use does not depend on the spectral type or absolute magnitude of the star. The ratio of M dwarfs that have flares varies with the spectral type, and this is included, but there is insufficient data (22 bursts on 5 stars) to determine whether the radio flux itself varies. The flares are not powered by the stellar luminosity but by magnetic activity. 

The parameters used in the Besançon simulation are slightly modified. The faint limit of the V band apparent magnitude is set to 23 because of the low luminosity of M dwarfs compared to OB stars. The spectral type is set to M. The latitude range is extended to $-$10 to +10 degree and because of their proximity, detected M dwarfs will not be as strongly focussed to the galactic plane.  Due to the large amount of M dwarfs in the galaxy, we only study a part of the sky ($0^{\circ} \leq \rm b \leq 10^{\circ}$) here.


\section{Results}

\subsection{Flux Distribution}
\begin{figure}[H]

\includegraphics[width=11cm]{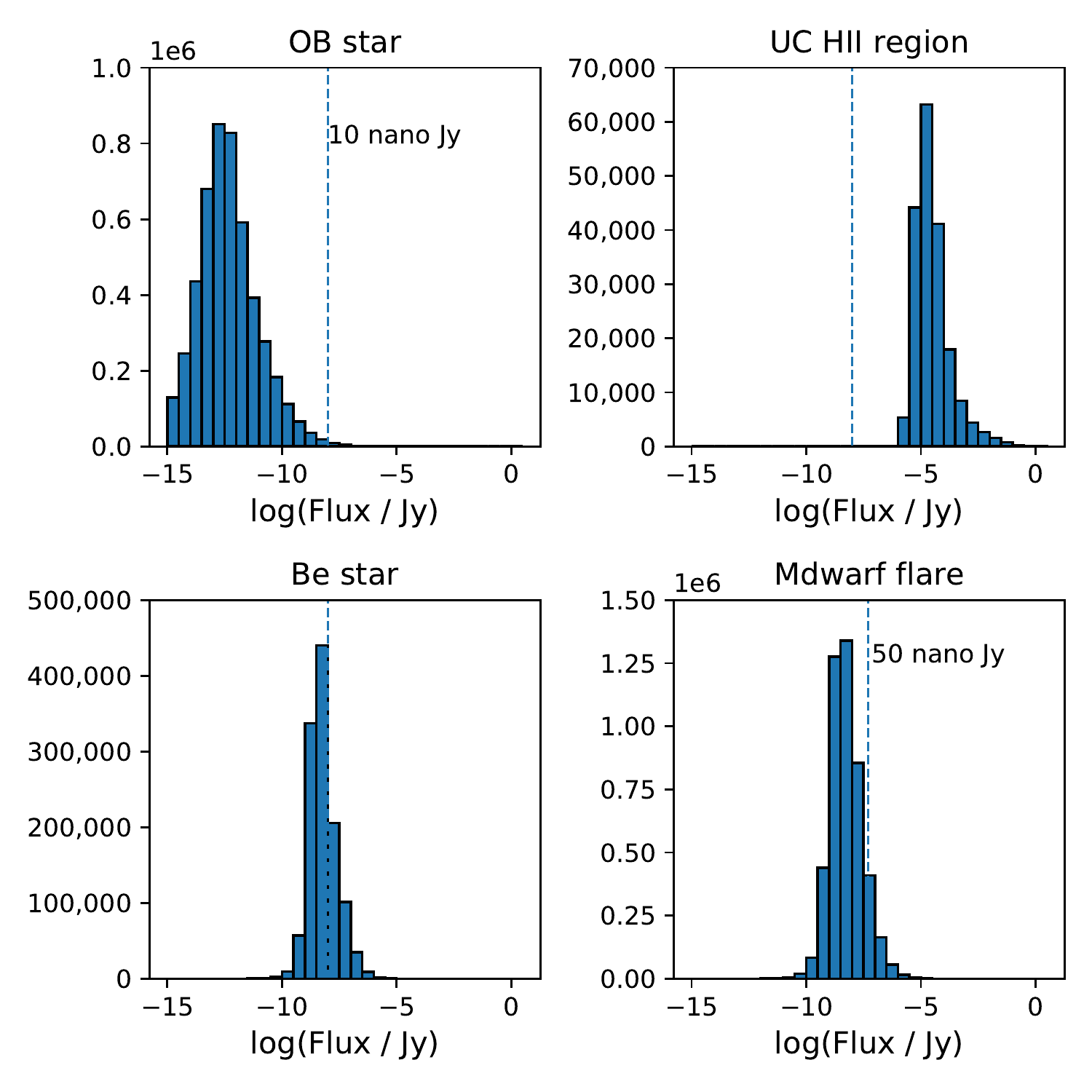}
\caption{The distribution of estimated flux of all our samples. The blue dashed line represents the assumed detection limit of the SKA.}
\label{fig:FluxAll}
\end{figure}

\begin{figure}[H]

\includegraphics[width=9cm]{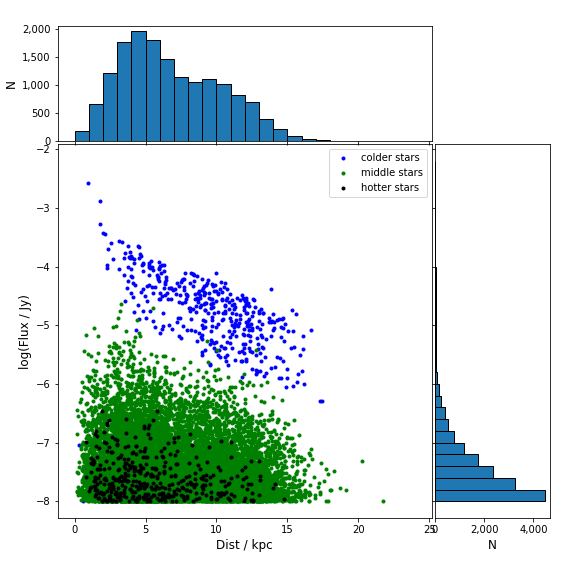}
\caption{The distribution of distance and estimated fluxes of detectable OB stars. The upper panel is for the distance, while the right is for the flux. The middle panel shows how the flux varies with distance. Black and green dots represent sources applied with  Equations~\eqref{eq:ml1} and~\eqref{eq:ml2}, respectively. Blue dots are sources with a temperature lower than 12,500 K.The flux gap in Figure~\ref{fig:OBDistr} rises from the bigger constant ($-$5.99) used in Equation~\eqref{eq:ml2}.}
\label{fig:OBDistr}
\end{figure}

\begin{figure}[H]

\includegraphics[width=9cm]{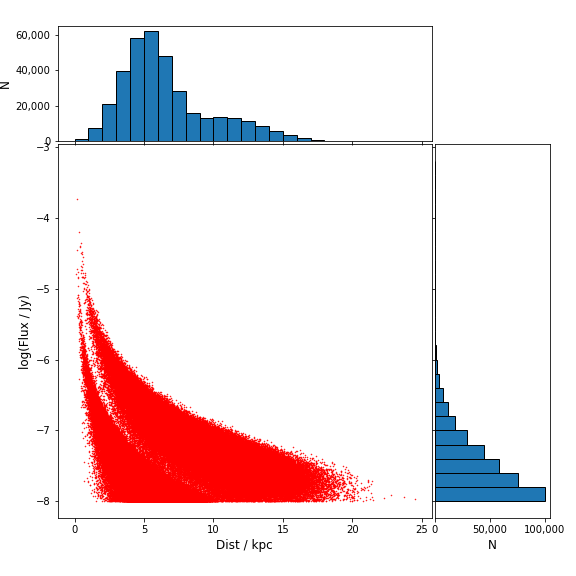}
\caption{The distribution of distance and estimated fluxes of detectable Be stars. The upper panel is for the distance, while the right is for the flux. The middle panel shows how the flux varies with distance. The apparent gap is caused by our flux prediction, which is bi-modal, as described in Section~\ref{sec:Be}. }
\label{fig:BeDistr}
\end{figure}

\begin{figure}[H]

\includegraphics[width=9cm]{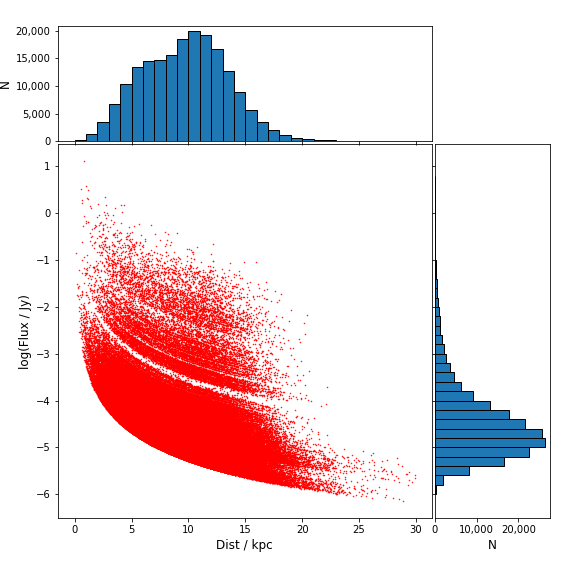}
\caption{The distribution of distance and estimated fluxes of detectable UC HII regions. The upper panel is for the distance, while the right is for the flux. The middle panel shows how the flux varies with distance. The gaps are due to the stellar temperature gridding in the Besançon model, combined with a very steep temperature-luminosity relation for the most massive main sequence stars.}
\label{fig:HIIDistr}
\end{figure}

\begin{figure}[H]

\includegraphics[width=9cm]{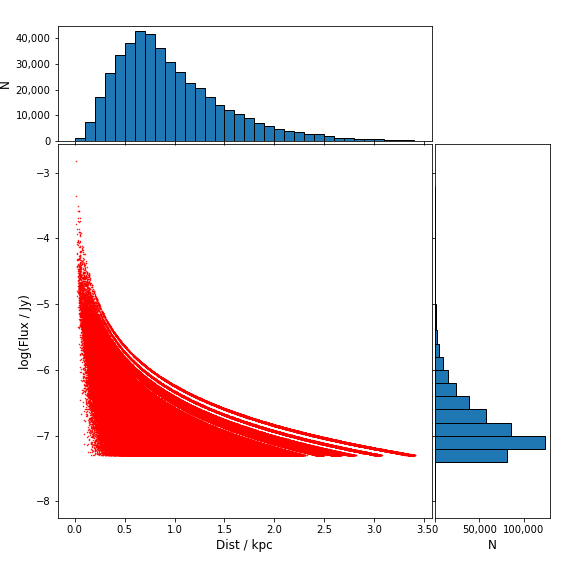}
\caption{The distribution of distance and estimated fluxes of detectable M dwarf flares. The upper panel is for the distance, while the right is for the flux. The middle panel shows how the flux varies with distance. The striping  is due to the binning used on the flare fluxes. The flare flux refers to the average flux over the 20-min flare.}
\label{fig:MfDistr}
\end{figure}

We estimated the flux of all our samples and compared to the assumed detection threshold of the full SKA. The results for all samples are shown in Figure~\ref{fig:FluxAll}, and only the detectable sources shown in the right panels of Figure~\ref{fig:OBDistr}--\ref{fig:MfDistr}. The numbers are for the selected sky areas of the samples. The detection limit is for 8~h of integration. For M dwarfs, it assumes a 20-min flare within the 8~h. As shown in the figures, only an extremely small fraction (about 0.3$\%$) of OB stars can be detected at 5~GHz. In contrast, the emissions from UC HII regions are strong enough that all of them are above the threshold, regardless of distance. This includes the more evolved HII regions around lower-mass stars. The numbers of potential detectable sources corresponding to a different threshold are summarized in Table \ref{table:fluxN}.


\begin{figure}[H]
\widefigure
\includegraphics[width=16cm]{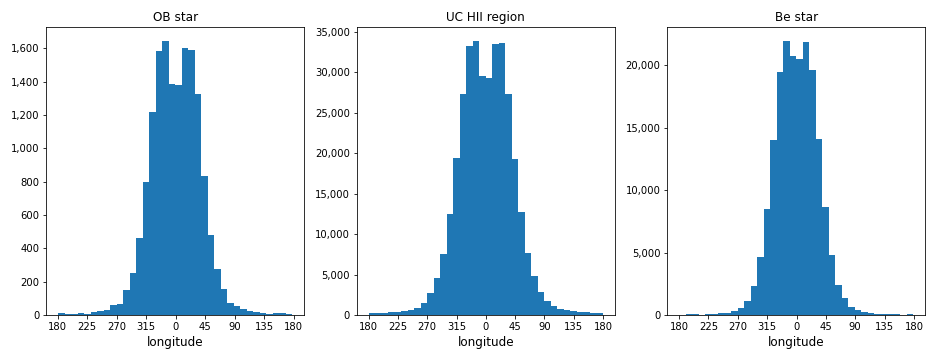}
\caption{The longitude 
 distribution of detectable OB, Be stars, and UC HII regions above~10 nano-Jy.}
\label{fig:lDistr}
\end{figure}

\begin{figure}[H]
\widefigure
\includegraphics[width=16cm]{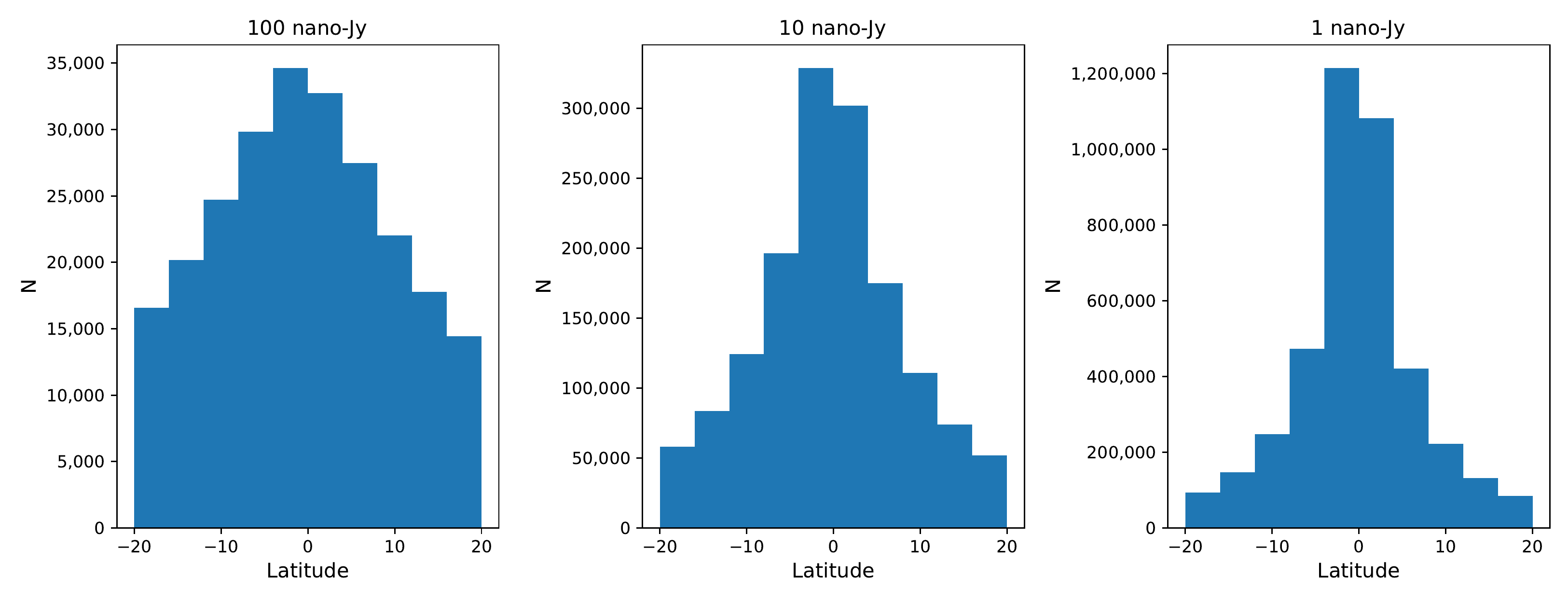}
\caption{The latitude distribution of M dwarf flares above different 8-hour detection limits (100 nano-Jy, 10 nano-Jy, and 1 nano-Jy), where the flare is assumed to last for 20 min. }
\label{fig:MbDistr}
\end{figure}

\begin{figure}[H]
\widefigure
\includegraphics[width=15cm]{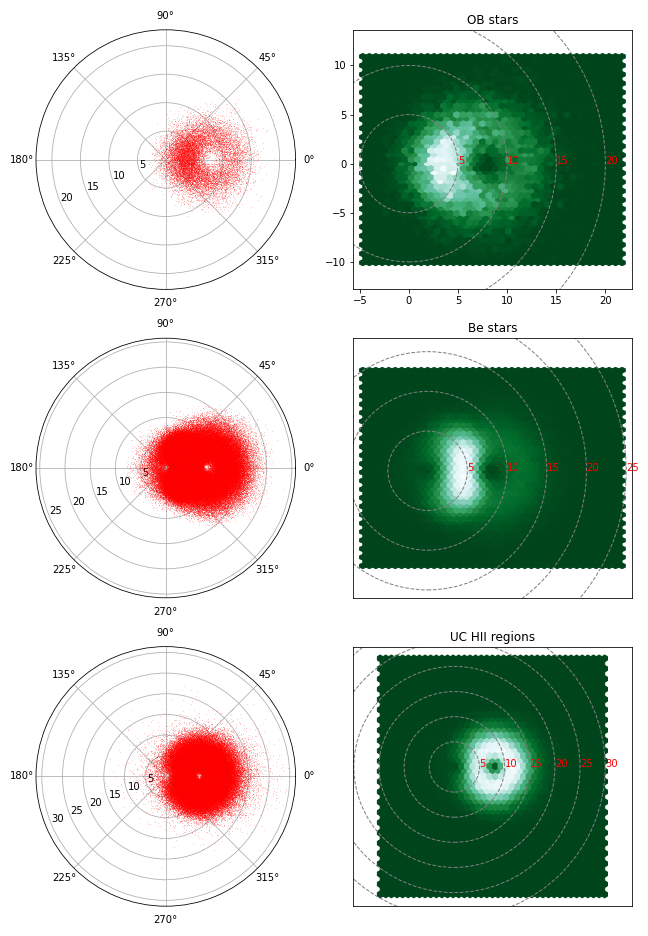}
\caption{The space distribution of detectable (above~10 nano-Jy in 8~h integration) OB, Be stars, and UC HII regions along the galactic plane. The first column is the source distribution in polar coordinates center at the sun. The radius represent distance to the sun in kpc. The second column is the kernel density distribution of the sources.}
\label{fig:SpaceDistr}
\end{figure}

\begin{figure}[H]
\widefigure
\includegraphics[width=15cm]{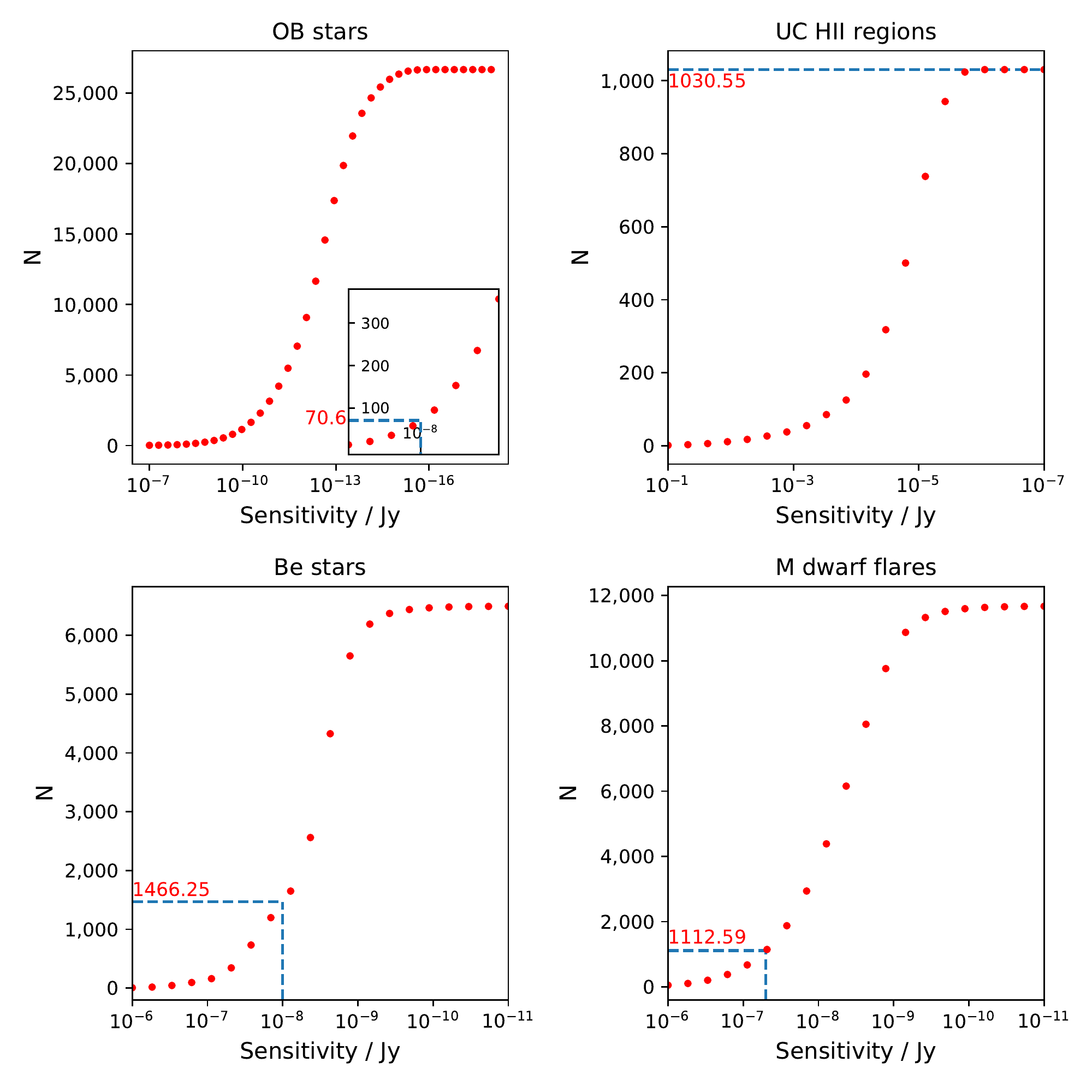}
\caption{Numbers of detectable sources per square degree toward the galactic center as a function of the detection thresholds. The blue dashed lines are the limit assumed in this work; the corresponding numbers of sources are given above it.}  
\label{fig:SourceDens}
\end{figure}

\begin{figure}[H]
\widefigure
\includegraphics[width=13cm]{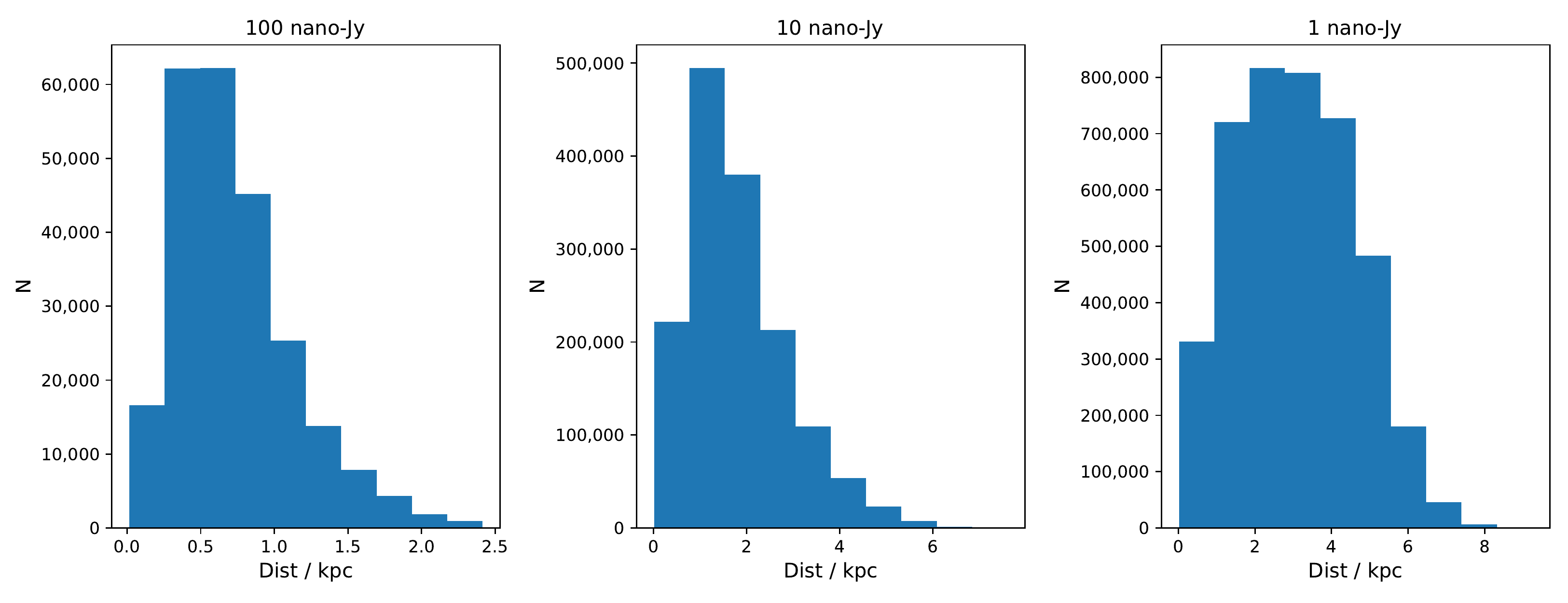}
\caption{The distance distribution of M dwarf flares above different detection limits (100 nano-Jy, 10 nano-Jy, and 1~nano-Jy).} 
\label{fig:MDisDisr}
\end{figure}

\vspace{-6pt}
\end{paracol}
\nointerlineskip
\begin{specialtable}[H]
\widetable
\caption{The number of detectable sources and fractions they account for corresponding to different detection limits.}
\label{table:fluxN}

\setlength{\cellWidtha}{\columnwidth/6-2\tabcolsep+.300in}
\setlength{\cellWidthb}{\columnwidth/6-2\tabcolsep-.100in}
\setlength{\cellWidthc}{\columnwidth/6-2\tabcolsep-00in}
\setlength{\cellWidthd}{\columnwidth/6-2\tabcolsep-.100in}
\setlength{\cellWidthe}{\columnwidth/6-2\tabcolsep-00in}
\setlength{\cellWidthf}{\columnwidth/6-2\tabcolsep-.100in}
\scalebox{1}[1]{\begin{tabularx}{\columnwidth}{>{\PreserveBackslash\centering}m{\cellWidtha}>{\PreserveBackslash\centering}m{\cellWidthb}>{\PreserveBackslash\centering}m{\cellWidthc}>{\PreserveBackslash\centering}m{\cellWidthd}>{\PreserveBackslash\centering}m{\cellWidthe}>{\PreserveBackslash\centering}m{\cellWidthf}}
\toprule
Flux Threshold (Jy)	& \boldmath{$10^{-6}$}	& \boldmath{$10^{-7}$}& $10^{-8}$	& \boldmath{$10^{-9}$}& \boldmath{$10^{-10}$}\\
\midrule
OB stars		& 598\,\, 0.012\%    & 2,651\,\, 0.054\%   & 15,676\,\, 0.319\%   & 70,577\,\, 1.437\%    & 248,121\,\, 5.051\% \\   
Be stars		& 1,966\,\, 0.16\%	 & 45,867\,\,  3.73\%   & 353,050\,\, 28.75\%  & 1,130,459\,\, 92.05\%  & 1,196,901\,\, 97.46\%\\
UC HII regions  & 189,895\,\, 99.99\% & 189,906\,\, 100.00\% & 189,906\,\, 100.00\% & 189,906\,\,  100.00\% & 189,906\,\,  100.00\%\\
M dwarf flares	& 19,366\,\,	0.40\%   & 234,009\,\, 4.88\%   & 1,485,377\,\, 30.66\% & 4,098,485\,\, 84.41\%  & 4,616,311\,\, 95.08\%\\
\bottomrule
\end{tabularx}}

\end{specialtable}
\begin{paracol}{2}
\switchcolumn

\vspace{-6pt}
\subsection{Sky Distribution}

The latitude distribution of our samples with detectable 5 GHz-continuum emission excluding  the M dwarfs is shown in Figure \ref{fig:lDistr}. As expected, there is a clear decreasing trend from the center to the anti-center. In Figure \ref{fig:MbDistr}, the latitude distribution of M dwarfs flares above the detection limits are presented. While they have a concentrated distribution around the plane, the numbers increase much faster at a low latitude when the detection threshold goes lower. Ergo, this means that a lot of faint, distant M dwarf flares hidden in the plane will be revealed when the sensitivity improves.

\clearpage
\end{paracol}
\nointerlineskip
\begin{paracol}{2}
\switchcolumn

\end{paracol}
\nointerlineskip
\begin{paracol}{2}
\switchcolumn

The two dimensional distribution of detectable OB, Be stars, and UC HII regions are shown in Figure \ref{fig:SpaceDistr}, along with their kernel density estimation (KDE) maps. As shown in the plot, the distribution of OB stars has a typical ring structure around the galactic center. The much higher density near the sun is due to the sensitivity limits; at a larger distance, many OB stars become too faint even for the SKA. The furthest OB detections reach to about 20 kpc. The hole around the galactic centre is because the Besançon model does not include the inner bulge. In the outer galactic plane, some fraction of stars is lost because of the warp.

The KDE map of Be stars has a very high density region near the sun. This results from our dualistic luminosity estimation mentioned in Section \ref{sec:Be}. The stars with a lower assumed luminosity are only detectable at close distances.  Those with a higher luminosity can be detected as far as 20 kpc away from the Sun. 

All of our samples of UC HII regions have a flux higher than the assumed detection limits. The distribution shows a symmetrical structure around the galactic center, basically covering all of the plane and reaching to almost 25 kpc away. Since we did not generate M dwarfs for all the sky, the sky coverage of M dwarf flares is not discussed here.

\clearpage
\end{paracol}
\nointerlineskip
\begin{paracol}{2}
\switchcolumn

We calculate the numbers of detectable sources per square degree for different detection limits toward the galactic center ($0\leq l \leq10^\circ$). These are plotted in Figure \ref{fig:SourceDens}. The expected number of sources per square degree towards other directions are listed in Table \ref{table:Nsquare}. The distance distribution of the four types of sources are shown in Figure \ref{fig:OBDistr}--\ref{fig:MfDistr}. In particular, we plot the distance distribution of detectable M dwarf flares corresponding to different sensitivities in Figure \ref{fig:MDisDisr}. As shown, the detection of flares is very sensitive to the flux limit. When the sensitivity is limited to 100 nano-Jy, the distance range is quickly cut to less than 2.5 kpc.

\end{paracol}
\nointerlineskip
\begin{paracol}{2}
\switchcolumn

\vspace{-6pt}

\begin{specialtable}[H]
\caption{The number of detectable sources (above 10 nJy) per square degree in different directions.}
\label{table:Nsquare}

\setlength{\cellWidtha}{\columnwidth/5-2\tabcolsep-00in}
\setlength{\cellWidthb}{\columnwidth/5-2\tabcolsep-00in}
\setlength{\cellWidthc}{\columnwidth/5-2\tabcolsep-00in}
\setlength{\cellWidthd}{\columnwidth/5-2\tabcolsep-00in}
\setlength{\cellWidthe}{\columnwidth/5-2\tabcolsep-00in}
\scalebox{1}[1]{\begin{tabularx}{\columnwidth}{>{\PreserveBackslash\centering}m{\cellWidtha}>{\PreserveBackslash\centering}m{\cellWidthb}>{\PreserveBackslash\centering}m{\cellWidthc}>{\PreserveBackslash\centering}m{\cellWidthd}>{\PreserveBackslash\centering}m{\cellWidthe}}
\toprule
Directions	& Center	& $\sim90^{\circ}$& Anti-Center	& $\sim270^{\circ}$\\
\midrule
OB stars		& 70.6\,\,0.265\%    & 3.05\,\,0.482\%   & 0.45\,\,0.616\%  & 3.15\,\,0.581\%\\   
Be stars		& 1,466.25\,\,22.06\%	 & 110.95\,\,70.42\%  & 15.8\,\,78.80\% & 102.6\,\,73.47\%\\
UC HII regions  & 1,030.55\,\,100.00\% & 25.15\,\,100.00\% & 3.2\,\,100.00\% & 23.0\,\,100.00\%\\
M dwarf flares	& 1,090.81\,\, 30.66\%	 & -  & - & -\\
\bottomrule
\end{tabularx}}

\end{specialtable}


\clearpage
\end{paracol}
\nointerlineskip
\begin{paracol}{2}
\switchcolumn

\vspace{-12pt}

\section{Discussion}

Our calculations show that the SKA at a sensitivity limit of 10nJy at 5~GHz, will detect all UC HII regions in the galaxy. In our calcluation, the total number is  nearly $2 \times 10^5$. The true number of UC HII regions is expected to be much lower as our number includes more evolved HII regions.  The sensitivity limit does not really come into it, as the faintest predicted flux in our model is 0.7 $\upmu$Jy, and the second faintest is just above 1 $\upmu$Jy. The number density in the galactic plane peaks at $10^3$ per square degree towards the centre, and drops to 3 per square degree in the anti-centre. For a beam size of the 15-m SKA dishes at 5~GHz of 0.3 degrees, or a beam area of 0.07 square degrees, we expect around 70 UC HII regions in the instantaneous field of view towards the galactic centre.

As the sources are very bright, an UC HII survey would be fast. Assuming 30-sec per pointing, the entire galactic plane visible from the SKA  within $\pm$1 deg could be covered in 4 days of observing. As a result of the very strong concentration towards the inner plane, a survey limited to $\pm$100 deg longitude would detect the large majority of UC HII regions.

The other types of sources studied are much fainter. To reach the 10 nJy sensitivity limit we assume will require 8~h of integration. At this limit, there are roughly 1500 Be stars per square degree in the densest area. If we want to detect 1000 Be stars, we need 10 instantaneous pointings, requiring 3 days of integration. We can achieve a similar number of detections  by observing a 10 times larger area to 100 nJy. The latter could be done in half a day but would miss the fainter population. In the same survey, we can also detect about 50 OB stars, because the density of detectable OB stars are only roughly 5 $\%$ of Be stars in the direction of Galactic center, and even lower in other directions.
 
For M dwarfs, the flux we estimated was for a 20-min integration, which is the assumed duration of the flare. (More precisely, we assume that the decay phase doubles the integrated flux over 8~h, compared to the 10-min probability distribution.) The typical probability of flaring at our sensitivity limit (assuming a distance of 1 kpc) is 1 to a few per cent. Over an 8-h integration, this means that we should expect a detected M dwarf to flare only once. The number of detected flares in 8~h is simply 24 times the number in 20 min. This gives roughly 4500 per square degree in the center direction. With 3 instantaneous pointings, we can obtain a sample of 1000 flares in approximately one day.

The M dwarfs are time variable. The SKA will detect them for some fraction of the observing time. This may affect the imaging process as not all observed UV points will see the same emission. 

In the outer plane, the Milky Way's warp affects the latitude distribution of the stars. It shifts the central plane by several degrees, depending on distance and longitude~\cite{Amores2017}. The warp is included in the Besançon model but is ignored for our latitude range from $-$1 to +1 degree. We may therefore miss some stars in  the outer plane. The missing fraction is negligible compared to all the selected stars, but it is important for longitudes between 90 and 270 degrees. We tested the distribution at longitude 120 degrees, and found that the predicted distribution peaks at $b \approx 1$ degree, at the edge of our selected range. We would miss up to half of all OB stars in this region, but would still include the densest regions.  Widening the latitude range to cover the warp at all distances would significantly increase the telescope time required for the plane survey. However, moving the selected area as a function of latitude could recover more stars in the outer galactic plane.
A total of 5 GHz is chosen for the reasons mentioned in Section \ref{sec:SKA}. However, not all stars have the steep thermal spectrum. UC HII regions have a flat spectra (if optically thin). Flare stars can be stronger at a lower frequency, peaking at 1--1.4~GHz~\cite{Villadsen2019}. For these objects, observations at a lower frequency may be advantageous, as this gives a much larger instantaneous field of view, while the OB and Be stars will become much~fainter.

\section{Conclusions}
Our results shows that radio emissions from OB, Be stars, UC HII regions, and M dwarfs flares could detectable with the full SKA achieving the 10 nJy sensitivity. Provided long enough integration times (8~h), a very large number of radio stars could be detected. All galactic UC HII regons could be detected. In the best conditions, roughly 70 OB stars, 1500~Be stars,  and 1000 M dwarf flares could be detected per square degree towards the center direction. The farthest objects reached to approximately 20 kpc from earth, except for M dwarf flares, whichweare limited to 3 kpc or less. There will be a significant number of flaring M dwarfs even outside the plane. The presence of faint, time variable sources need to be taken into account when devising observational strategies low frequency surveys. We also present estimates in Table \ref{table:fluxN} and Figure \ref{fig:SourceDens} for higher flux limits, in case the sensitivity of the SKA cannot reach 10 nano-Jy in the future. In this case, integration times longer than 8~h may be needed.

The SKA1 could reach 300 nJy in 8~h. This will allow a full survey of galactic HII regions, but for stellar emissions only the brightest and closest stars can be detected. The detection rate of OB stars, Be stars, and M dwarfs with the SKA1 will be 10\% of the SKA2 sample, and will not trace the fainter population. The full survey requires the full SKA.

\vspace{6pt} 



\authorcontributions{ The concept was by A.Z., the methodology, software and analysis was done by B.Y., the draft was written by B.Y. and edited by A.Z. and B.J., supervision of the project was by A.Z. and B.J. All auhtors have read and agreed to the published version of the manuscript.}

\funding{This research was funded by the UK Science and Technology Facility Council (STFC) under the GCRF---Researcher Exchange Programme  and by the China Scholarship Council (CSC) under grant no. 201706040320. This work was supported by National Key R$\&$D Program of China no. 2019YFA0405503, the CSST Milky Way Survey on Dust and Extinction Project and NSFC 11533002.}

\institutionalreview{Not applicable.}

\informedconsent{Not applicable.}

\acknowledgments{This research greatly benefited from discussions with Keith Grainge. }

\conflictsofinterest{The authors declare no conflict of interest. The funders had no role in the design of the study; in the collection, analyses, or interpretation of data; in the writing of the manuscript, or in the decision to publish the results.}




\end{paracol}
\reftitle{References}






\end{document}